\documentclass[aps,prl,twocolumn,superscriptaddress]{revtex4-1}  %
\usepackage{graphicx}  %
\usepackage{dcolumn}   %
\usepackage{bm}        %
\usepackage{amssymb}   %

\begin{document}

\widetext
\leftline{Version xx as of \today}

\title{Topological solitons as addressable phase bits in a driven laser}

\author{Bruno Garbin}
\affiliation{Universit\'e de Nice - CNRS UMR 7335, Institut Non Lin\'eaire de Nice, 1361 route des lucioles 06560 Valbonne, France}
\author{Julien Javaloyes} 
\affiliation{Departament de F\'isica, Universitat de les Illes Baleares, C/ Valldemossa km 7.5, 07122 Mallorca, Spain}
\author{Giovanna Tissoni}
\affiliation{Universit\'e de Nice - CNRS UMR 7335, Institut Non Lin\'eaire de Nice, 1361 route des lucioles 06560 Valbonne, France}
\author{St\'ephane Barland}
\email{stephane.barland@inln.cnrs.fr}
\affiliation{Universit\'e de Nice - CNRS UMR 7335, Institut Non Lin\'eaire de Nice, 1361 route des lucioles 06560 Valbonne, France}

\date{\today}

\begin{abstract}

Optical localized states are usually defined as self-localized bistable packets of light which exist as independently controllable optical intensity pulses either in the longitudinal or transverse dimension of nonlinear optical systems. Here we provide the first experimental and analytical demonstration of the existence of longitudinal localized states which exist fundamentally in the phase of laser light. These robust and versatile phase bits can be individually nucleated and canceled in an injection-locked semiconductor laser operated in a neuron-like excitable regime and submitted to delayed feedback. The demonstration of their control opens the way to their use as phase information units in next generation coherent communication systems. We analyze our observations in terms of a generic model which confirms the topological nature of the phase bits and discloses their formal but profound analogy with Sine-Gordon solitons.

\end{abstract}

\maketitle

\section{introduction}

Dissipative solitons have been observed in many nonlinear optical devices, both in the dimension transverse to light propagation and along the propagation dimension. Numerous examples have been found in nonlinear optical resonators with coherent forcing (they often receive the name of spatial or temporal cavity solitons) \cite{solitette,SichM.2011,temporalcs,herr2014temporal,AdvAtMolOpt57,barbay2011cavity} but also in laser systems \cite{Grelu2012,tanguy:013907,genevet08,elsass2010}. In the latter case, the phase of the electric field is free to evolve in the course of time and a paradigmatic model is the cubic-quintic Ginzburg-Landau equation \cite{Grelu2012,akhmediev,akhmediev2008dissipative,fauve90}. On the contrary, the presence of an external field leads to the formation of dissipative solitons whose phase is locked to this external forcing in the second case, whose paradigmatic and first model is the Lugiato-Lefever equation \cite{PhysRevLett.58.2209}. In spite of this important difference \cite{nphoton.2010.1}, the dissipative solitons observed in all of these systems are in most cases explained by a double compensation of dispersion (or diffraction) by Kerr nonlinearity and losses and gain \cite{nphoton.2010.1,akhmediev,akhmediev2008dissipative,rosanov_spatial_hysteresis}. 

In this work, we report the first experimental observation of a new kind of dissipative
optical localized states which are fundamentally phase objects. Contrary to well-known dissipative optical solitons (phase locked or not), the localized states we generate and control can not exist in either of the paradigmatic models and their existence does not result from the usual double balance, but from the phase space topology of the system which supports them. Nevertheless, since they are attractors of a nonlinear and dissipative systems, they share with usual dissipative solitons their discrete and robust character. These properties, together with their nature of optical phase objects, confers them exceptional properties as phase bits ($\Phi$-bits) for coherent optical communications.

Our experiment is based on a semiconductor laser with coherent optical injection in a regime called \textit{excitable} \cite{coulletexcitwaves,PhysRevLett.98}, with the addition of a delayed feedback loop. A system is said to be \textit{excitable}  when any perturbation which is sufficient to overcome the \textit{excitability threshold} elicits an always identical response, whose details does not depend on the perturbation. Paradigmatic examples include neural or cardiac cells \cite{winfree1980the,keener2008mathematical}.

In this configuration and without feedback, the phase of the semiconductor laser
is stably locked to the external forcing, except when responding to
a sufficiently large external perturbation \cite{turconi2013control} which triggers a $2\pi$ phase rotation after
which the system locks again to the external forcing. We use the delayed
feedback as a spatial degree of freedom \cite{giacomelli1996relationship,giacomelli2012coarsening,larger2013virtual,PhysRevLett.112.103901} in which multiple independently
addressable $\Phi$-bits can be stored indefinitely. These structures
are embedded in a homogeneously locked background field and therefore
draw their robustness from their topological nature. Besides their
fundamental novelty, these $\Phi$-bits open new perspectives for
the generation of multidimensional optical localized states. Finally,
the possibility to nucleate and annihilate these states brings the
plasticity and robustness of optical localized states to the realm
of phase information processing in coherent optical communication
networks \cite{slavik2010all,radic2010optical,kakande2011multilevel}.

\section{Results}
\begin{figure*}
	\centering \includegraphics[width=0.65\textwidth]{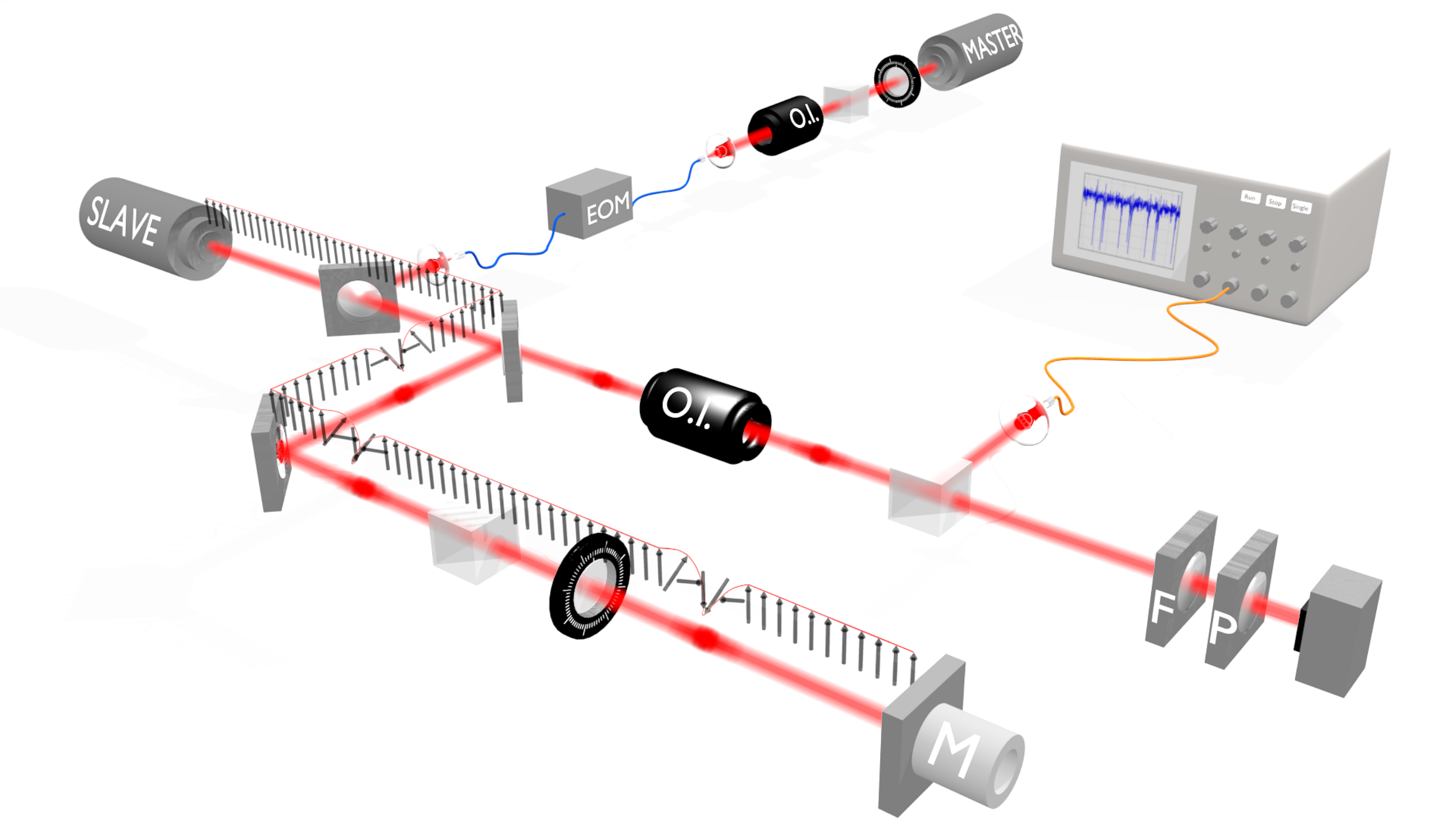} \caption{Experimental setup. The slave laser is submitted to optical injection by the master laser. Unidirectional coupling is ensured by an optical isolator (OI) and phase perturbations can be applied to the slave laser via the fiber-coupled electro-optic phase modulator (EOM). The arrows show a schematic view in the $\Re(E),\Im(E)$ plane of several $\Phi$-bits stored in the feedback loop.}
  \label{fig:setup}
\end{figure*}

The experimental arrangement is shown on Fig. \ref{fig:setup}. It
consists of a single transverse and longitudinal mode semiconductor
laser (called ``slave laser'') under the  action
of coherent external forcing and delayed optical feedback. In order
to obtain the desired phase space topology, the slave laser is biased
at very high pump value (six to eight times the standalone lasing threshold, emitted power about $500 \mu$W) and the master laser is tuned such that the detuning
between both lasers is close to 5~GHz. The power of the injection beam
($2$ to $3\mu$W depending on realizations) is set such that the slave laser is in a stable stationary state
in which its phase is locked to the external forcing. The injection
beam detuning and power, together with the bias current of the slave
laser are chosen such that the system is excitable
\cite{turconi2013control}. This parameter regime allows
to essentially confine the dynamics in phase space on a circle whose radius is set by the emitted power and which contains two
fixed points, one stable and one unstable, representing two different
values of the relative phase of the slave laser with respect to the
forcing. 

Perturbations can be applied to the system via a phase modulator. In order to confer the system
an analogue of spatial degree of freedom, we place it inside an optical feedback
loop. This is achieved by directing part of the emitted intensity towards a polarizer, quarter wave plate and high reflectivity mirror
which constitute a feedback loop of 0.5\% to 1\% reflectivity. The mirror is initially placed at
a distance of about 30~cm and its sub-wavelength positioning
is set via a piezo-electric actuator. 

\begin{figure}
\centering \includegraphics[width=0.45\textwidth]{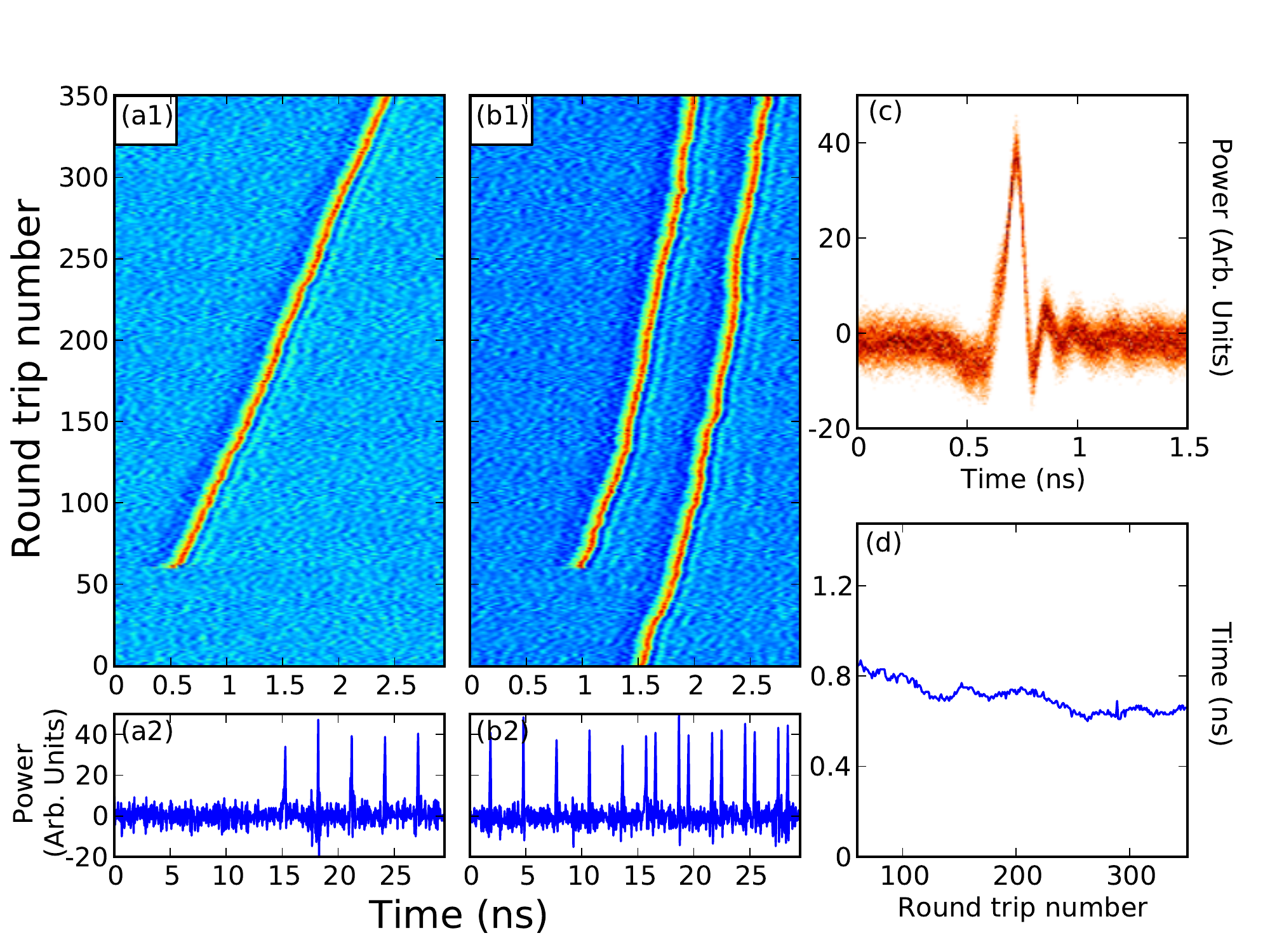} 
\caption{Nucleation of two $\Phi$-bits in space-time representation. On panel
	(a1), a phase perturbation is applied (black arrow) while the system is in a stable
stationary locked state. Following that perturbation, a pulse is nucleated
and repeats with a periodicity close to the feedback delay time, as
shown on panel (a2). The space-time representation is chosen such
that the pulse is almost stationary (see text). After some time a
phase perturbation is applied again on the system (black arrow), with the first
$\Phi$-bit already stored. The two $\Phi$-bits now propagate in
the feedback loop, without perturbing each other (panels (b1) and
(b2). Panel (c) shows a superposition of a single $\Phi$-bit waveform over 350 roundtrips.  The very well defined shape indicates the attractor nature of the $\Phi$-bit. The ringing following the pulse is attributed to the detection setup. Panel (d) shows the evolution of the distance between the
two pulses in the course of time.}
  \label{fig:indep}
\end{figure}

The detection apparatus comprises a Fabry-Perot interferometer
for spectral monitoring and a 9~GHz
photo-detector whose output is further amplified by a 14~GHz AC-coupled
amplifier whose output is acquired via a 12.5~GHz bandwidth (100GS/s) real
time oscilloscope. The detection is isolated
from the experiment by a 30~dB optical isolator to prevent
spurious reflections towards the slave laser.

In absence of the optical feedback, the system responds to suitable phase perturbations \cite{turconi2013control} by emitting a single excitable pulse which is a homoclinic orbit in phase space \cite{coulletexcitwaves}. This pulse consists of the relative phase between slave and master laser completing a full circle before settling again to the initial value \cite{kelleher2009excitable,turconi2013control}, in excellent analogy with an over-damped pendulum submitted to a fluid torque or excitable particles in an optical torque wrench \cite{pedaci2010}. Slow (200~ps) or wrongly oriented phase variations do not trigger the excitable response \cite{turconi2013control}.
 
This response is detected as a small pulse in the emitted power (10 to 25\% of DC value depending on parameters). 

In presence of very weak optical feedback with adequate phase, 
we observe that a pulse with essentially identical characteristics
(amplitude and duration) is emitted, but also \textit{regenerated}
after a delay very close to the delay time. This observation is reported
on panels (a1) and (a2) of figure \ref{fig:indep}. Panel (a2) shows
the time trace as measured by the detection apparatus, but the trajectory
of the pulse in the feedback cavity is best observed in the co-moving
reference frame (see methods), as shown on panel (a1). On this graph,
the horizontal axis is a space-like coordinate $x$ (in nanoseconds) chosen such that the pulse
is almost stationary and the vertical coordinate $\xi$ (in units of $x$) corresponds to the temporal
evolution of the system over many round-trips. At this point, the
possibility of using the spatial degree of freedom of the system to
store information becomes clear. We show the first demonstration of
this on figure \ref{fig:indep}, panels (b1) and (b2). The pulse created
previously on panel (a1) is now at position $x=1.5$~ns at time $\xi=0$
and slowly drifting to the right. At round-trip number $\xi=60$ and space
position about $x=1$~ns, a phase perturbation is applied on the
injected beam via the phase modulator. As previously, this perturbation
triggers a pulse, which is regenerated periodically due to the optical feedback. The mutual independence of these
pulses demonstrates their nature of ``localized structures''. Of
the utmost importance, we underline that all the data shown on figure
\ref{fig:indep} consists in real time data. This enables the measure
of the speed of the pulses in the feedback loop (see Methods). 

We analyze the evolution of the shape of a single $\Phi$-bit in the
course of time by plotting on panel (c) of figure \ref{fig:indep} the superposition of a single $\Phi$-bit over 350 roundtrips. The very well defined shape indicates the attractor nature of the $\Phi$-bit.

Finally, we show on panel (d) of Fig. \ref{fig:indep} the
evolution of the distance between the $\Phi$-bits over successive round-trips. Again, the slow evolution of the distance
is a confirmation of the interpretation of the $\Phi$-bits in terms
of mutually independent localized states and not simply an harmonic
solution of the fundamental period set by the feedback loop.

\begin{figure}
\centering \includegraphics[width=0.5\textwidth]{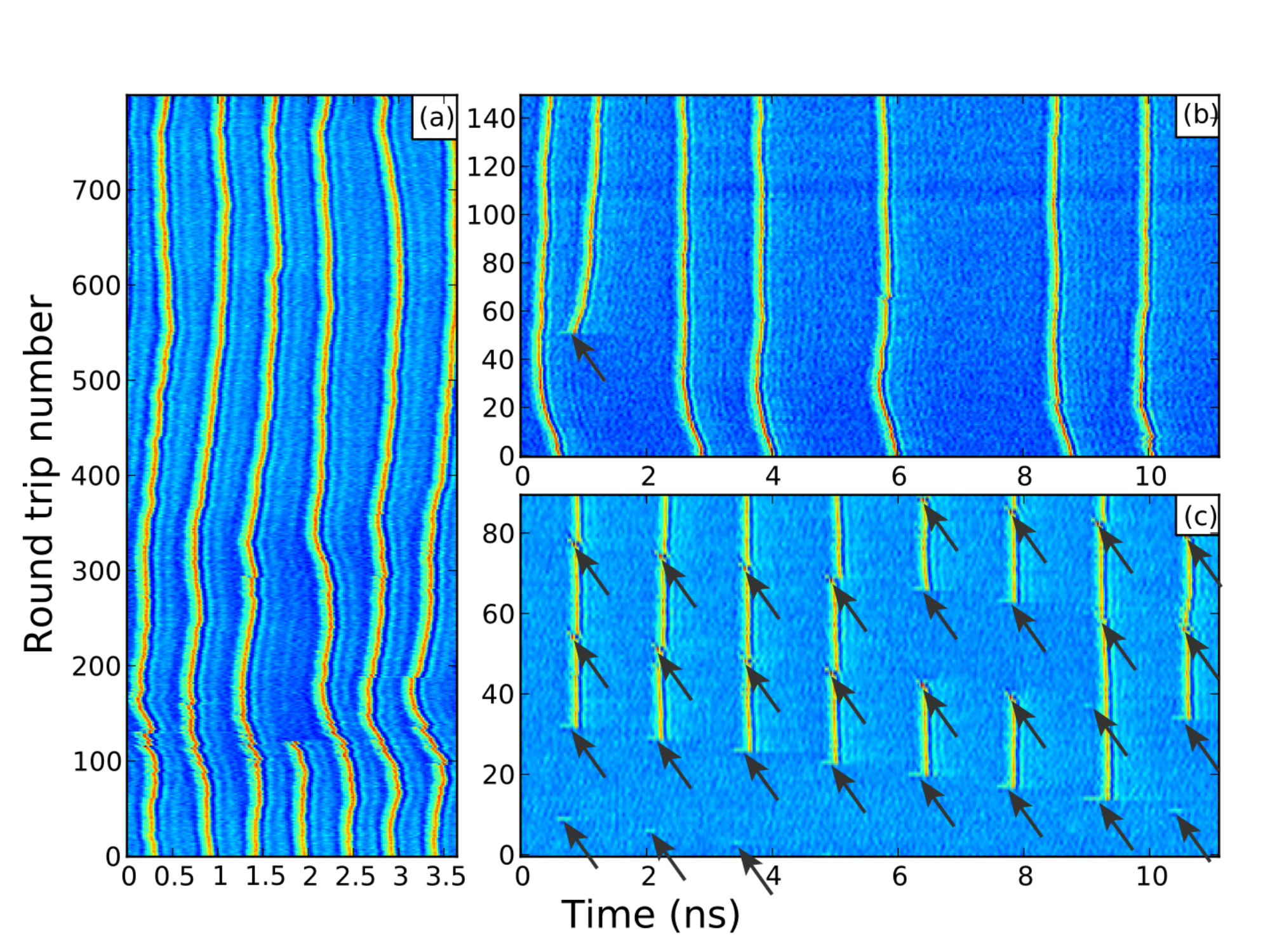}
\caption{Control of $\Phi$-bits. Panel a): seven $\Phi$-bits initially coexist in the feedback loop. The six remaining bits reorganize in the course of time after cancellation of the central one. Panel b): 6 $\Phi$-bits are present in the feedback loop. A perturbation (black arrow) is applied and nucleates a seventh $\Phi$-bit. The two nearest neighbor repel each other until they reach sufficient time separation (790~ps)\, but the others existing $\Phi$-bits are not affected. Panel c): Many perturbations (black arrows) are applied successively in time and nucleate several $\Phi$-bits starting from the homogeneous state. Existing $\Phi$-bits can be canceled if they are hit by a phase perturbation\, as shown at approximate coordinates (6~;40) and (8~;40). In this case, a nonmonotonous sequence of 13 different 8-bits integers has been stored. The same perturbation parameters are used for nucleation and cancellation. Note that the feedback loop is three times larger in b) and c) than in a).}
  \label{fig:control}
\end{figure}

As becomes evident from figure \ref{fig:indep}, independent topological
localized structures can be juxtaposed only in a space which is large
enough, otherwise interactions may set in. In the latter case all the peaks would move all together. This is illustrated on figure \ref{fig:control}, panel (a). In this case, seven $\Phi$-bits have been stored in the memory. At about round-trip 100, the
central structure vanishes spontaneously due to electronic noise in
the bias current control system which also perturbs the other $\Phi$-bits.
At that point, we observe that upon cancellation of one of the structures,
the other structures smoothly reorganize, the fourth one drifting
to the left, later followed by the fifth one. We notice
that this reorganization does not consist of an abrupt
reconfiguration of the ensemble, but a rather a slow motion of
each pulse confirming their weak interaction. The fluctuating distances
indicate the degrees of freedom associated to the separations between
pulses, which we analyze theoretically as the last part of our results.

\begin{figure*}
\centering \includegraphics[width=0.99\textwidth]{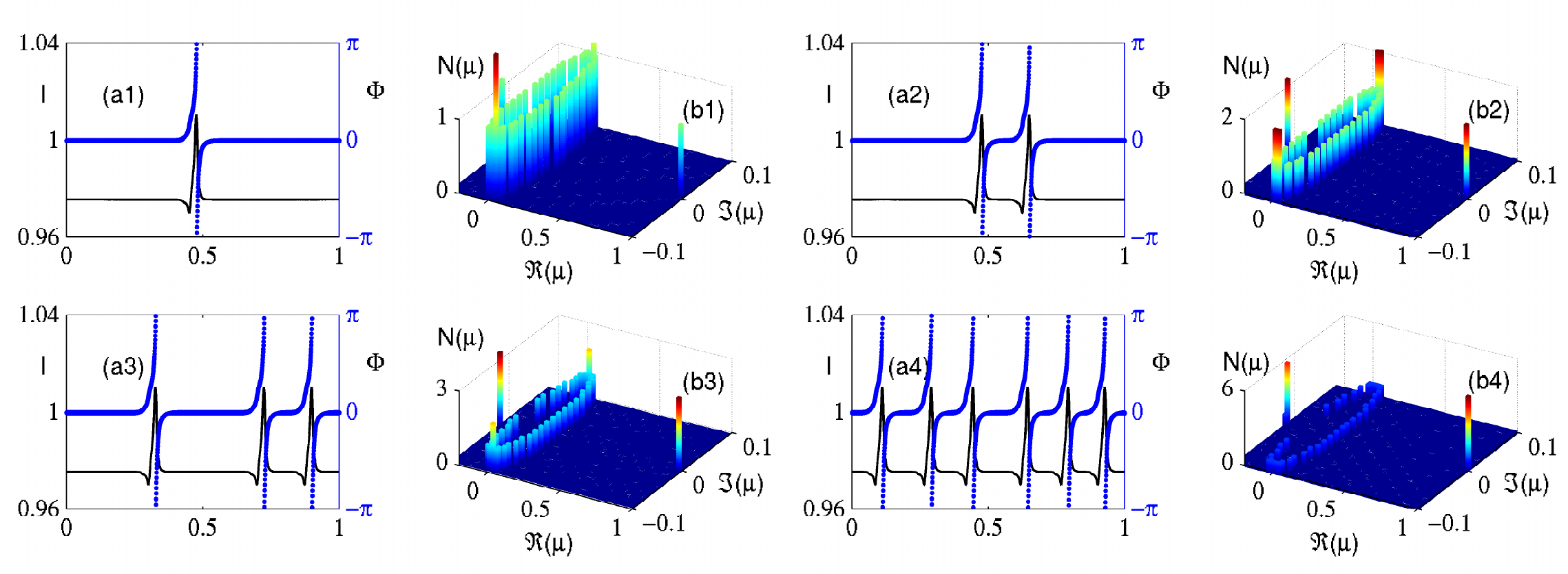}
\caption{Analysis of  Ginzburg-Landau model with forcing and delayed feedback. Temporal trace (a) for the output intensity $I=|A|^{2}$ and phase
$\Phi=\arg\left(A\right)$ and histogram $N\left(\mu\right)$ of the
Floquet multipliers $\mu$ (b) in the cases of 1,2,3 and 6 $\Phi$-bits.
One notices that the number of neutral modes located in the vicinity
of $\mu=1$ increases linearly with the number of $\Phi$-bits. The
parameters are $\alpha=3$, $\Delta=0.1$, $Y=2.9\times10^{-2}$,
$\eta=5\times10^{-3}$, $\Omega=0$, $k=1$ and $\tau=2000$. }
  \label{fig:theory}
\end{figure*}

In order to minimize the interaction between structures, the feedback
loop must be much larger than many times the typical interaction distance. The experimental setup is therefore
modified to increase the available space and the feedback mirror is
set at a distance of about $1.65$~m. We show on figure \ref{fig:control}
the addressing of many $\Phi$-bits and show that they can exist
at different distances from each other. We start (panel b))with six $\Phi$-bits
previously nucleated and coexisting in the cavity. At round-trip 50,
a perturbation is applied which nucleates a new bit. This new structure
is nucleated with a temporal separation of 570~ps
from another existing structure. Consequently, the new structure is
repelled from the nearest neighbor in a way that is strongly reminiscent
of the ultra-weak force between temporal cavity solitons \cite{jang2013ultraweak,prati2013solitons}
but in the present case we can  attribute this interaction
to the refractory time following excitable pulses \cite{izhikevich2006dynamical}. This
interaction lasts about 50 round-trips. At this point, the structures
have repelled to a temporal separation of about 790~ps
after which no strong repulsion is observed and the distance is essentially
randomly fluctuating. The absence of interaction between the pulse
is clear on top of panel (b), in which the structures are
separated by arbitrary distances.

Finally, we demonstrate the nucleation \textit{and annihilation}
of structures on panel (c). To this aim, we repetitively apply phase
perturbations at a repetition rate of 31.3~MHz, which trigger excitable pulses in different points of the space-time diagram, starting
from an empty cavity (bottom of panel (c)). Some perturbations (at
coordinates $(x,\xi)=(3.5;5),(2;8),(0.8;10),(10.5;12)$) are
hardly visible and did not nucleate structures due to fluctuations blurring the excitability threshold \cite{pedaci2010,turconi2013control}. On the contrary, seven successful nucleations then take
place and equally spaced pulses separated by approximately 1.5~ns
in the feedback loop are generated. The eighth perturbation then appears
again on the right of the co-moving reference frame and nucleates
an eighth $\Phi$-bit. At this point, we have demonstrated the coexistence
of nine different solutions holding from 0 to 8 $\Phi$-bits. The
ninth perturbation arrives very close to an existing structure but
does not perturb it strongly. On the contrary, the following two perturbations
reach the system close enough to existing $\Phi$-bits to annihilate
them. Subsequent phase kicks do not perfectly superimpose with existing
structures and therefore only perturb them without destroying them.
From round-trips 40 to 60, the system has been switched to another
existing state in which only 6 $\Phi$-bits are present in the cavity.
This state differs from the other 6-$\Phi$-bits state which exists
around round-trip 28 by the configuration of the bits. This configuration can contain information. At round-trip
62 and 64 new structures are nucleated again in the space which was previously
freed by the cancellation achieved close to round-trip 40. It is of
course difficult to annihilate these temporal localized states since
applying a perturbation to them requires accurately aiming 
in time. In this case, we have achieved it by perturbing in a repetitive
way at a period which is not too far from a multiple of the round-trip
time. The result shown on figure \ref{fig:control} constitutes the
first demonstration of optical annihilation of temporal localized
states, and also a demonstration of the switching between many states
including from 0 to 8 $\Phi$-bits in different configurations.

The experimental observations can be interpreted within the generic
framework of the injected Ginzburg-Landau equation extended to the
influence of a delayed feedback. The equation governing
the evolution of the optical field $E$ reads

\begin{equation}
\frac{dE}{dt} = \left(1+i\alpha\right)\left(1-\vert E\vert^{2}\right)E+Y+i\Delta E+\eta e^{-i\Omega}E_{\tau}.\label{eq:CA}
\end{equation}

In equation (\ref{eq:CA}), time has been scaled to the photon lifetime
in the cavity times the excess pumping above threshold, while $\alpha$
and $\Delta=\omega_L-\omega_Y$ stand for the linewidth enhancement factor and the detuning
between the solitary laser frequency $\omega_L$ and the injected
field $\omega_Y$, respectively. The optical feedback $E_{\tau}=E\left(t-\tau\right)$
with time delay $\tau$ has an amplitude and a phase denoted $\eta$
and $\Omega$, respectively. The field amplitudes $E$
and $Y$ have been rescaled to the one of the solitary laser and $Y\in \mathbb{R}$ by definiteness. Although such
a model may not represent the whole complexity of the system, it can
be derived rigorously from standard laser equations close to threshold \cite{JMP-PRE-03}.
 It is also one of the simplest paradigm that is able
to capture the physics of the problem and the important ingredients
of the phase space geometry. Importantly, it allows us to connect with the 
experimental results by noting that the output of the VCSEL consists in a superposition of
the emitted and reflected fields which reads $A=E-kY$ with
$k=\left(1-r_{1}r_{2}\right)[\left(1-r_{1}\right)\left(1+r_{2}\right)]^{-1}\sim1$
with $r_{1,2}>0$ the resonant reflectivities in amplitude of the top and bottom distributed Bragg Reflectors. The value of k is deduced by solving exactly the wave propagation within the linear empty regions of the VCSEL as in \cite{MB-JQE-05} (see supplementary material).

Our results are summarized in Fig.~\ref{fig:theory}(a1)-(a4) in
which we represent the (residual) temporal evolution of the intensity
as well as the phase dynamics of the output field in good agreement
with the experimental results. We present as well in Fig.~\ref{fig:theory}(b1)-(b4)
the stability of such periodic solutions. The stability information
was obtained via a partial diagonalization of the Monodromy operator
(see method). The monodromy Matrix $\mathcal{M}$ describes the evolution
of a perturbation after a full period. A solution is stable if all the 
eigenvalues of $\mathcal{M}$, the so-called Floquet multipliers, correspond
to damped motion, i.e. $\mathrm{Max}|\mu|\le1$. Due to the infinite
number of Floquet multipliers in a delayed differential equation,
we represented in Fig \ref{fig:theory}(b) a histogram of the multipliers
$N\left(\mu\right)$ for the sake of clarity. In any time invariant
dynamical system, a periodic solution must present a Floquet multiplier
equal to the unity. Often termed the trivial multiplier,
it merely represents the translational temporal invariance and physically
amounts to ``shifting''  the solution without having to pay
an ``energetic'' cost. The details of the temporal waveform (e.g.
with several peaks over one period) are irrelevant. Such trivial multiplier
is visible for the single $\Phi-$bit solution. However, we demonstrate
in Fig.~\ref{fig:theory}(b2)-(b4) that the solutions with $N$ $\Phi-$bits
are much more than a single multi-peaked periodic solution: they
present not one but $N$ multipliers clustered around $\mu=1$. This
fact has a profound impact on the dynamics as it implies that the $N$
$\Phi-$bits solution possesses $N$ neutral modes, which is actually
what one would expect for independent $\Phi-$bits. We analyzed the
eigenvectors associated with the various neutral multipliers and found
that they correspond to relative translations of each of the $\Phi-$bits
thereby confirming their independence.

For weak injection and detuning, one can, by applying a multiple time
scale analysis to Eq.~(\ref{eq:CA}) reduce the problem even further to a single delayed
equation for the phase that reads
\begin{eqnarray}
\frac{d\theta}{dt'} & = & \Delta'-\sin\theta+\chi\sin\left(\theta_{\tau'}-\theta-\psi\right),\label{eq:phi}
\end{eqnarray}
with $\theta=\Phi+\arctan\alpha$, $\psi=\Omega+\arctan\alpha$, $\chi=\eta/Y$ and $\left(1/\Delta',t',\tau'\right)=\left(1/\Delta,t,\tau\right)/\left(Y\sqrt{1+\alpha^{2}}\right)$
confirming that the dynamics consists essentially in a phase phenomena. 
Interestingly, the same reduction to Eq.~(\ref{eq:phi}) is possible far from threshold 
from a full Class-B laser model with nonlinear gain compression in good correspondence 
with the experimental conditions.

Such phase model allows us to exploit the strong link between delayed
systems and spatio-temporal dynamics that was established in \cite{giacomelli1996relationship}.
By applying a multiple time scale analysis to Eq.~(\ref{eq:phi}) similar
to the one in \cite{giacomelli1996relationship} (see supplementary
material) we formally reduce Eq.~(\ref{eq:phi}) to the modified Sine-Gordon
equation 
\begin{eqnarray}
\frac{\partial\theta}{\partial\xi} & = & \sin\bar{\theta}-\sin\theta+\frac{\partial^{2}\theta}{\partial x^{2}}+\tan\psi\left(\frac{\partial\theta}{\partial x}\right)^{2},\label{eq:Sine-Gordon}
\end{eqnarray}
with $\sin\bar{\theta}=\Delta'-\sin\psi$, $x$ the pseudo-space variable
and $\xi$ a slow temporal variable. When $\sin\bar{\theta}=\tan\psi=0$, analytical
$2\pi$ homoclinic orbits corresponding to kink solutions
of this equation are known as 
\begin{eqnarray}
\theta\left(x\right) & = & 4\arctan\exp x.\label{eq:kink}
\end{eqnarray}
In the general case, it was shown in \cite{coulletexcitwaves} that these
homoclinic loops verify the Melnikov condition and therefore are robust
in the limit $\sin\bar{\theta}\ll1$ and $\tan\psi\ll1$.
They therefore persists in our case and we found that they seem to agree well
with the numerical solutions of Eq.~(\ref{eq:phi}) even far from such perturbative limit. 

\section{discussion}

We demonstrated experimentally the optical control of independent 
phase bits in an excitable system with time delayed feedback. We evidenced
the nucleation of structures and analyzed their basic interaction
which is repulsive, consistent with the underlying refractory
tail of excitable pulses. The duration of each $\Phi$-bit
is close to 100~ps  with typical interaction
length less than 800 ps. Pulse of 70~ps duration (bandwidth limited) were also observed. These values are
basically set by the detuning between the slave laser and the external
forcing. Since the $\Phi$-bits are essentially
phase objects, the semiconductor medium dynamics should not have a strong impact and we expect
that much shorter pulses could be realized with obvious benefits for information capacity. Finally the nucleation of the $\Phi$-bits was performed via phase perturbations, but wavelength and intensity-phase conversions can be expected since excitable pulses in injection locked semiconductor lasers can also be incoherently triggered \cite{Garbin:14}.

The theoretical analysis which has been presented, based on a 
generic model of a laser system with the addition of forcing and delayed
feedback confirmed the paradigm that is required for the generation
of $\Phi$-bits, namely the presence of a saddle node on a circle
bifurcation with the addition of a delay term. Due to the simplicity
of the model we have been able to analyze numerically the neutral
modes associated to the relative translation of the $\Phi$-bits with
respect to each other, which confirms their nature of localized states
beyond the coexistence of multiple solutions, which was also observed.
In addition, we exploited the strong relation between delayed and
spatio-temporal systems which allowed us to interpret the temporal
$\Phi-$bits as robust homoclinic kinks of a modified Sine-Gordon equation.
Due to the genericity of our results, we believe that next-generation photonic sources such as quantum cascade \cite{Faist22041994} or polariton \cite{schneider2013electrically} lasers may support $\Phi$-bits with identical dynamical origin and features and very attractive physical properties.

In conclusion we have presented the first observation of temporal
localized structures which are topological in nature and thus essentially
phase objects. These phase bits are attractors of an out of equilibrium
system and they can be controlled independently of each other. As
such, they can provide not only information storage, but also pulse reshaping and discrimination functionalities
for the phase data which is the basis of coherent optical communications
networks.

\section{Methods}

\subsection{Experimental details}

\subsubsection{Setup}
The laser used in the experiment is a Vertical Cavity Surface Emitting Laser (ULM980-03-TN-S46). It emits close to 980~nm in a single longitudinal and transverse mode and is linearly polarized up to 1.8~mA with coherent emission threshold at 0.2~mA. It is driven by a $1~\mu$A resolution power supply and actively temperature stabilized. The output of the Slave Laser is collected by 4.5~mm focal length collimator. A half wave plate situated right after the collimator allows to align the polarization of the Slave Laser with the vertical axis. A 10\% beam splitter is placed in front of it to serve as input for the Master Laser beam. The master laser is a tunable edge emitting laser with external grating in Littrow configuration. The amount of injected power can be precisely set via rotation of a half-wave plate placed between the master laser and a polarizer. The detuning between slave and master laser is set as $\Delta=\omega_L - \omega_Y > 0$.

Phase perturbations can be applied to the system by applying voltage pulses to a fiber coupled 10~GHz Lithium Niobate phase modulator which is driven by a 100~ps (10\%-90\%)rise time pulse generator. The feedback cavity is \textit{not} actively stabilized but is placed inside two layers of enclosure in order to avoid alterations of the feedback phase condition due to air circulation.

\subsubsection{Co-moving reference frame}
In order to clearly visualize the data, we process it such that the observer is in the co-moving reference frame of the $\Phi$-bits. This is achieved by acquiring a \textit{single} long enough time trace (in this case up to $10^7$ points) and splitting this unique, perfectly synchronized time series into segments whose length corresponds to the time taken by a pulse to go to and back from the feedback mirror. The segments are then stacked on top of each other, so as to constitute a space-time diagram often used to analyze data in time-delayed dynamical systems \cite{giacomelli1996relationship, giacomelli2012coarsening, larger2013virtual,PhysRevLett.112.103901}. In order for the pulses to be perfectly stationary, the length of the segments $x$ must be $x = \tau + \delta$ where $\tau$ is the delay time. The additional delay $\delta$ results from the drift term of delayed dynamical systems \cite{giacomelli1996relationship,giacomelli2012coarsening} and from the fact that the $\Phi$-bits, which are phase objects, may not propagate exactly at the group velocity which is used to define $\tau$. 
This procedure allows us to detect minute changes in $\delta$ which enables to observe for instance the synchronous change of direction of motion of all $\Phi$-bits in Fig. \ref{fig:control} panel b). Although we cannot separate the two terms contributing to $\delta$, its total value in the parameter regimes used here has been found to be of the order of 12\% of $\tau$. On the other hand, $\tau$ was independently estimated by observing beat notes in non stationary regimes and the modal structure of such aperiodic complex regimes could also slightly deviate from the exact time of flight in the external cavity.

\subsection*{Theory}

\subsubsection{Numerical simulations}

The delayed differential equation given by the equation (\ref{eq:CA})
was numerically integrated with a fourth-order Runge-Kutta method
with constant step size ($\delta t=10^{-2}$)~\cite{NR-BOOK}. The
delayed contribution in Eq.~(\ref{eq:CA}) demands a special care.
To advance the solution with a step $h$ from $t_{n}=n\delta t$ to
$t_{n+1}$, the Runge-Kutta algorithm requires the values of $E\left(t-\tau\right)$
at intermediate points $t_{n+1/2}$. These are not known and must
be interpolated from past values with an order of approximation consistent
with that of the algorithm of integration. Therefore, besides keeping
memory of the past values of $E$ we also retain the past values of
the time derivative $\dot{E}\left(t\right)$. Such a method allows
building a third order Hermite polynomial approximation for $E(t)$
between the time $\left(t_{n}-\tau\right)$ and $\left(t_{n+1}-\tau\right)$.
By evaluating this interpolating polynomial at $\left(t_{n+1/2}-\tau\right)$,
we ensure an overall fourth order accuracy.

\subsubsection{Stability analysis}

The linear stability analysis of the periodic solutions of Eq.~(\ref{eq:CA})
was performed via the reconstruction of the monodromy operator $\mathcal{M}$.
Although a priori infinite dimensional, the operator reduces to a
matrix of size $\tau/\delta t$ due to the discrete sampling incurred
by the constant step-size numerical algorithm. Taking one point of
the periodic orbit, we insert a small perturbation in all the degrees
of freedom as represented by the mesh points in the delay time and
let the system evolve over one period. The deviation of the end point
from the unperturbed orbit yields a column of the operator $\mathcal{M}$.
This method bears some similarity to the one developed in \cite{PJB-OE-11}
but here over a period and not a single time step. Due to the large
size $\dim\left(\mathcal{M}\right)\sim10^{5}$, the eigenvalues and
eigenvectors can not be calculated from a complete decomposition using
for instance the QR method\cite{NR-BOOK}. Instead, we exploited the
sparsity of $\mathcal{M}$ and relied on the so-called Implicitly
Restarted Arnoldi Method \cite{IRAM} searching for the eigenvalues
of largest modulus.

\section{Acknowledgements}

JJ acknowledges financial support from the Ram\'on y Cajal program, 
project RANGER (TEC2012-38864-C03-01) and from Direcci\'o 
General de Recerca de les Illes Balears co-funded by the European Union 
FEDER funds as well as CNRS for supporting a visit at INLN where part of 
his work was developed. BG, GT and SB acknowledge support from R\'egion 
Provence Alpes C\^ote d'Azur through grant number DEB 12-1538. 
SB acknowledges funding from Agence Nationale de la Recherche 
through grant number ANR-12-JS04-0002-01.

\section{Author contributions}
B. Garbin performed the experimental characterization under the 
supervision of S. Barland as well as the numerical simulations under the 
direction of G. Tissoni. J. Javaloyes developed the theoretical analysis 
and the numerical characterization of the Floquet exponents. S. Barland 
wrote the manuscript assisted by G. Tissoni and J. Javaloyes.

\section{Conflict of interest}
The authors declare no competing financial interests.

\clearpage

\section*{Supplementary material}

\subsection*{Class-A to phase equation}

We start with the class-A laser equation
\begin{equation}
\frac{dE}{dt}  \\=\  \left(1+i\alpha\right)\left(1-\vert E\vert^{2}\right)E+Y+i\Delta A+\eta e^{-i\Omega}E_{\tau},\label{eq:CA-1}
\end{equation}
 By defining $E=\rho e^{i\phi}$ we get 
\begin{eqnarray}
\frac{d\rho}{dt} &\! =\! & \left(1-\rho^{2}\right)\rho+Y\cos\phi+\eta\rho_{\tau}\cos\left(\Delta\phi-\Omega\right),\\
\frac{d\phi}{dt} &\! =\! & \alpha\left(1-\rho^{2}\right)+\Delta-\frac{Y}{\rho}\sin\phi+\eta\frac{\rho_{\tau}}{\rho}\sin\left(\Delta\phi-\Omega\right).
\end{eqnarray}
where we defined $\Delta\phi=\phi_{\tau}-\phi$. Assuming that $Y$
, $\eta$ and $\Delta$ are of order $\varepsilon$, and inserting
a multiple time scale expansion as
\begin{eqnarray}
\rho\left(t_1,\varepsilon t_2\right) & = & 1+\varepsilon r\left(t_1,\varepsilon t_2\right)+\mathcal{O}\left(\varepsilon^{2}\right),\\
\phi\left(t_1,\varepsilon t_2\right) & = & \phi\left(t_1,\varepsilon t_2\right),\\
\frac{d}{dt} & = & \frac{\partial}{\partial t_1}+\varepsilon\frac{\partial}{\partial t_2},
\end{eqnarray}
we find that the phase does not depend on the fast time
scale, i.e. $\partial\phi/\partial t_1=0$, while at first order
we obtain
\begin{eqnarray}
\frac{\partial r}{\partial t_1} & = & -2r+Y\cos\phi+\eta\cos\left(\Delta\phi-\Omega\right),\\
\frac{\partial\phi}{\partial t_2} & = & -2\alpha r+\Delta-Y\sin\phi+\eta\sin\left(\Delta\phi-\Omega\right).
\end{eqnarray}
We can now perform the adiabatic elimination of the fast variable
$r$ on the time scale $t_1$ and find that it is slaved to the slow
variations of $\phi$ on the time scale $t_2$ as
\begin{eqnarray}
2r & = & Y\cos\phi+\eta\cos\left(\Delta\phi-\Omega\right).
\end{eqnarray}

Upon replacing the expression of $r$ we get 
\begin{eqnarray}
\frac{1}{\sqrt{1+\alpha^{2}}}\frac{d\phi}{dt_2} & = & \frac{\Delta}{\sqrt{1+\alpha^{2}}}-Y\sin\left(\phi+u\right)\\
 & + & \eta\sin\left(\Delta\phi-\Omega-u\right),
\end{eqnarray}
where we defined $u=\arctan\alpha$. Finally, rescaling the time and the delay as $(t',\tau')=(t,\tau) Y\sqrt{1+\alpha^{2}}$
which incidentally redefines the detuning and the feedback rate as
$\Delta'=\Delta/\left(Y\sqrt{1+\alpha^{2}}\right)$ and $\chi=\eta/Y$ yields
\begin{eqnarray}
\frac{d\phi}{dt'} & = & \Delta'-\sin\left(u+\phi\right)+\chi\sin\left(\phi_{\tau'}-\phi-\Omega-u\right).
\end{eqnarray}

A last possible simplification amounts in redefining the phase $\theta=u+\phi$
and introducing $\psi=\Omega+u$ reducing the problem to 
\begin{eqnarray}
\frac{d\theta}{dt'} & = & \Delta'-\sin\theta+\chi\sin\left(\theta_{\tau'}-\theta-\psi\right).\label{eq:phi-1}
\end{eqnarray}
The phase model contains only three parameters, $\Delta'$ the strength
of the detuning with respect to the injection, the feedback rate and its phase, 
the delay being mostly irrelevant as long as it
is large compared to the internal time scale of Eq.~(\ref{eq:phi-1}).
The saddle-node steady states $\bar{\theta}$ and $\tilde{\theta}$
are defined as $\bar{\theta}=\arcsin\left(\Delta'-\chi\sin\psi\right)$
and $\tilde{\theta}=\pi-\bar{\theta}$.

\subsection{Sine-Gordon spatio-temporal equation}

Assuming that the solution is almost of period $\tau'$ we get that $\theta_{\tau'}-\theta\ll1$
hence, we can expand Eq.~(\ref{eq:phi-1}) as
\begin{eqnarray}
\frac{d\theta}{dt'} & = &\left(\Delta'-\sin\theta\right)-\chi\sin\psi+\chi\cos\psi\left(\theta_{\tau'}-\theta\right)\\
 & + & \frac{\chi\sin\psi}{2}\left(\theta_{\tau'}-\theta\right)^{2}+\cdots
\end{eqnarray}

Similarly to the analysis close to an Andronov-Hopf bifurcation presented in \cite{giacomelli1996relationship}, 
we introduce two time scales $u_{1}=t'$ and $u_{2}=\varepsilon^{2}t'$, the chain rule being $d_{t'}\rightarrow\partial_{u_{1}}+\varepsilon^{2}\partial_{u_{2}}$. 
The central point in our analysis lies in the way we expand the delayed contribution as $\theta_{\tau'}=\theta\left( u_{1}+\varepsilon \upsilon,\varepsilon^{2}u_{2}-\varepsilon^{2}\tau'\right)$.
Doing so, we assume that the solution evolves slowly from one round-trip toward the next, which is described by the variations on the slow time $u_2$, 
up to a small drift term $\varepsilon \upsilon$ whose contribution is accounted for in the variable $u_1$. Expanding the delayed term up to second order leads to 

\begin{eqnarray}
\theta_{\tau'}-\theta & \sim & \left(\varepsilon \upsilon \frac{\partial\theta}{\partial u_{1}}+\frac{(\varepsilon \upsilon)^{2}}{2}\frac{\partial^{2}\theta}{\partial u_{1}^{2}}-\varepsilon^{2}\tau'\frac{\partial\theta}{\partial u_{2}}\right),\\
\left(\theta_{\tau'}-\theta\right)^{2} & \sim & (\varepsilon \upsilon)^{2}\left(\frac{\partial\theta}{\partial u_{1}}\right)^{2}+\mathcal{O}\left(\varepsilon^{3}\right),
\end{eqnarray}
which eventually yields the modified Sine-Gordon equation with a tilt as well as a quadratic velocity contribution 
\begin{eqnarray}
\frac{\partial\theta}{\partial\xi} & = & \sin\bar{\theta}-\sin\theta+\frac{\partial^{2}\theta}{\partial x^{2}}+\tan\psi\left(\frac{\partial\theta}{\partial x}\right)^{2}.\label{eq:SG}
\end{eqnarray}
In Eq.~(\ref{eq:SG}), we removed the drift velocity induced by the feedback term by identifying that $\varepsilon \upsilon=[\chi \cos(\psi)]^{-1}$. 
In addition, we rescaled the slow time by defining $\xi=u_{2}\varepsilon^{-2}/\left(1+\tau'\chi\cos\psi\right)$
and scaled the pseudo-space as $x=u_1 \sqrt{2\chi\cos\psi}$.

\subsection{Superposition coefficient with the incoming field}

We will use the methodology developed in \cite{MB-JQE-05} which amount
in solving exactly the wave propagation within the linear empty regions
of the VCSEL. Our starting point is the scalar Maxwell equation in
the temporal Fourier representation 
\begin{eqnarray}
\left(\partial_{z}^{2}+\nabla_{\perp}^{2}\right)\mathcal{E}\left(\omega,\vec{r}\right)+\frac{\omega^{2}}{c^{2}}n^{2}\mathcal{E}\left(\omega,\vec{r}\right) & = & \frac{-\omega^{2}}{c^{2}\varepsilon_{0}}\mathcal{P}\left(\omega,\vec{r}\right).
\end{eqnarray}

Assuming that the transverse profile of index guiding defines the
modal structure of the resonator, we can use the effective index method
and reduce the problem to the longitudinal dimension only. We obtain
for $\mathcal{E}=E\left(\omega,z\right)\Theta\left(r_{\perp}\right)$
with $\Theta\left(r_{\perp}\right)$ the transverse mode with effective
index $n_{e}$ 
\begin{eqnarray}
\left(\partial_{z}^{2}+\frac{\omega^{2}}{c^{2}}n_{e}^{2}\right)E\left(\omega,z\right) & = & -\frac{\omega^{2}}{c^{2}\varepsilon_{0}}P\left(\omega\right)\delta\left(z-l\right),
\end{eqnarray}
 where we assumed that the Quantum-wells (QWs) are located at $z=l$
in a device of length $L$ starting at $z=0$. In the region where
there is no polarization the solution is simply a superposition of left and right propagating waves 
\begin{eqnarray}
E\left(\omega,z\right) & = & E_{+}\left(\omega\right)e^{iqz}+E_{-}\left(\omega\right)e^{-iqz}
\end{eqnarray}
 with the longitudinal wave-vector $q=n_{e}\omega/c$. Integrating
over the QW width in $z=l$ yields
\begin{eqnarray}
\partial_{z}E\left(\omega,l^{+}\right)-\partial_{z}E\left(\omega,l^{-}\right) & = & -\frac{\omega^{2}}{\varepsilon_{0}c^{2}}P\left(\omega\right).
\end{eqnarray}

We next decompose the field into forward backward components in the
left and right part of the cavity as 
\begin{eqnarray}
L\left(z\right) & = & L_{+}e^{iqz}+L_{-}e^{-iqz},\\
R\left(z\right) & = & R_{+}e^{iqz}+R_{-}e^{-iqz},
\end{eqnarray}
yielding the four boundary conditions that consist in the two reflections 
and transmissions at the top and bottom mirrors as well as the continuity of the field and
the discontinuity of its derivative imposed by the presence on the plane of the QWs in n $z=l$ as 
\begin{eqnarray}
r_{1}L_{-}+t_{1}^{'}\tilde{Y} & = & L_{+},\\
r_{2}R_{+}e^{iqL} & = & R_{-}e^{-iqL},\\
L\left(l\right) & = & R\left(l\right)=\mathcal{E}\left(\omega,l\right),\\
\partial_{z}R\left(l\right)-\partial_{z}L\left(l\right) & = & -\frac{\omega^{2}}{\varepsilon_{0}c^{2}}P,
\end{eqnarray}
with $r_{1}$ and $r_{2}$ the top (emitting) and bottom reflectivities,
respectively. The primes correspond to reflection and transmission
processes starting outside of the cavity. The amplitude of the external
field impinging on the device is denoted $\tilde{Y}$. Expressing all the
fields as a function of the field on the QWs $E\left(\omega,l\right)=E$
we find 
\begin{eqnarray}
F_{1}\left(q\right)E+F_{2}\left(q\right)\tilde{Y} & = & \frac{-\omega^{2}}{iq\varepsilon_{0}c^{2}}P,
\end{eqnarray}
with
\begin{eqnarray}
F_{1} & = & \frac{2\left(1-r_{1}r_{2}e^{2iqL}\right)}{\left(1+r_{1}e^{2iql}\right)\left(1+r_{2}e^{2iq\left(L-l\right)}\right)},\\
F_{2} & = & \frac{-2t_{1}^{'}e^{iql}}{1+r_{1}e^{2iql}}.
\end{eqnarray}

The modes of the VCSEL cavity are defined by a minima of the function
$|F_{1}|$, i.e. $1-r_{1}r_{2}e^{2iqL}\sim0$ that is to say a $L=m\lambda$
cavity. Because we demand gain, we also want a maxima of the modal confinement factor $\Gamma=\left(1+r_{1}e^{2iql}\right)\left(1+r_{2}e^{2iq\left(L-l\right)}\right)$
that is to say $\Gamma\sim\left(1+r_{1}\right)\left(1+r_{2}\right)$.
In other words, we impose the presence of an antinode of the field on the QW which selects
incidentally only an even number of $\frac{\lambda}{2}$ oscillations. 
Since the numerator of $F_1$ is the strongly varying function we only expand $1-r_{1}r_{2}e^{\left(\cdots\right)}$ 
and neglect the small variations of $F_{2}$. Imposing that $q$ is close to resonance $q_{0}$ for which
$\exp\left(2iq_{0}L\right)=\exp\left(2iq_{0}\left(l=\frac{L}{2}\right)\right)=1$,
we obtain 
\begin{eqnarray}
1-r_{1}r_{2}e^{2iqL} & \sim & \left(1-r_{1}r_{2}\right)-r_{1}r_{2}\left(q-q_{0}\right)2iL_{e}\nonumber \\
t_{1}^{'}e^{iql}\left(1+r_{2}e^{2iq\left(L-l\right)}\right) & \sim & t_{1}^{'}\left(-1\right)^{m}\left(1+r_{2}\right)+\cdots
\end{eqnarray}
where the length $L_{e}=L-\frac{i}{2}\left(\partial_{Q}\ln r_{2}+\partial_{Q}\ln r_{1}\right)$
takes into account the frequency dependence of $r_{j}\left(\omega\right)$,
i.e. the frequency dependent penetration depth into the DBRs. Expanding $q$ around $q_{0}=n\omega_{0}/c$
saying that $\omega\rightarrow\omega_{0}+i\partial_{t}$ yields the
final field evolution equation as 
\begin{eqnarray}
\tau\frac{dE}{dt} & = & \frac{i\omega_{0}\left(1+r_{1}\right)\left(1+r_{2}\right)}{2n\varepsilon_{0}cr_{1}r_{2}}P-\frac{1-r_{1}r_{2}}{r_{1}r_{2}}E\nonumber \\
 & + & t_{1}^{'}\frac{1+r_{2}}{r_{1}r_{2}}\left(-1\right)^{m}\tilde{Y}.
\end{eqnarray}

As a last step, we must evaluate the field at the laser output $A$
as a combination of the reflection of the injected beam $r_{1}^{'}\tilde{Y}$
as well as transmission of the left intracavity propagating field
$L_{-}$, i.e. 
\begin{eqnarray}
A & = & t_{1}L_{-}+r_{1}^{'}\tilde{Y}\\
 & = & E\frac{t_{1}e^{iql}}{1+r_{1}e^{2iql}}-\tilde{Y}\frac{r_{1}+e^{2iql}}{1+r_{1}e^{2iql}}
\end{eqnarray}
where we used the Stokes relation $tt'-rr'=1$ and $r'=-r$ to simplify the last equation. 
Around resonance such expression simplifies into 
\begin{eqnarray}
A & = & E\frac{t_{1}\left(-1\right)^{m}}{1+r_{1}}-\tilde{Y}
\end{eqnarray}
The $\left(-1\right)^{m}$ is irrelevant in the sense that the QW
field experiences also an injection field with a $\left(-1\right)^{m}$
meaning that the superposition is always in anti-phase and can be
absorbed into a redefinition of $E\rightarrow-E$, hence in the following
we will assume $m$ even. Introducing the adiabatic expression for the gain 
as $P=g(\alpha-i)(N-N^t)$ as well as the carrier equation of evolution, 
scaling time to the photon lifetime of device
$\kappa=\frac{1-r_{1}r_{2}}{r_{1}r_{2}\tau}$ and redefining the injected
field as $Y=h\tilde{Y}$  with $ h = t_1^{'} (1+r_2)/(r_1 r_2) $ yields 
\begin{eqnarray}
\dot{E} & = & \left[\left(1+i\alpha\right)D-1\right]E+Y,\label{eq:Laser1}\\
T\dot{D} & = & J-\left(1+\vert E\vert^{2}\right)D,\label{eq:Laser2}\\
A & = & \frac{t_1}{1+r_1}\left(E- k Y \right)\sim E-k Y.\label{eq:Laser3}
\end{eqnarray}

with the following definition of k as
\begin{eqnarray}
k & = & \frac{1-r_{1}r_{2}}{\left(1-r_{1}\right)\left(1+r_{2}\right)},\label{eq:k_cagas}
\end{eqnarray}
where we used again the Stokes relation, defined $T=\kappa/\gamma_{\parallel}$ and 
normalized the carriers as $D=\Gamma g \omega_0 (2n\epsilon_0)^{-1}(N-N^t)$.
\end{document}